\documentclass[letterpaper]{article}

\newif\ifALIFE
\ALIFEtrue
\ifALIFE
	\usepackage{natbib,alifeconf}  
\else
	\usepackage{natbib}
	\usepackage{times}
	\usepackage{graphicx}

\twocolumn
\labelwidth\leftmargini\advance\labelwidth-\labelsep \labelsep 5pt

\date{May 2020}

\fi

\usepackage{array}
\usepackage[hyphens]{url}
\usepackage{hyperref}

\title{Lenia and Expanded Universe}
\author{Bert Wang-Chak Chan \\
\mbox{}\\
Hong Kong \\
albert.chak@gmail.com}

\begin{document}
\maketitle

\begin{abstract}
We report experimental extensions of Lenia, a continuous cellular automata family capable of producing lifelike self-organizing autonomous patterns. The rule of Lenia was generalized into higher dimensions, multiple kernels, and multiple channels. The final architecture approaches what can be seen as a recurrent convolutional neural network. Using semi-automatic search e.g. genetic algorithm, we discovered new phenomena like polyhedral symmetries, individuality, self-replication, emission, growth by ingestion, and saw the emergence of ``virtual eukaryotes'' that possess internal division of labor and type differentiation. We discuss the results in the contexts of biology, artificial life, and artificial intelligence.
\end{abstract}

\section{Introduction}

The study of cellular automata (CA) is one of the major branches in artificial life and complex systems research. CAs were invented by John von Neumann and Stanislaw Ulam \citep{VonNeumann1951,Ulam1962}, then popularized by John H. Conway's Game of Life (GoL) \citep{Gardner1970} and Stephen Wolfram's elementary cellular automata (ECA) \citep{Wolfram1983}. On the one hand, research on CAs led to proofs of Turing completeness and therefore the capability for universal computation in CAs, e.g. GoL and ECA Rule 110 \citep{Rendell2002,Cook2004}. On the other hand, CAs were utilized to model complex systems, generate patterns, and produce computer art.

One line of investigation involves attempts to construct long-range or continuous CAs, search for and study self-organizing autonomous patterns, or {\em solitons}. These attempts include CAPOW \citep{Rucker1999}, Larger-than-Life \citep{Evans2001}, RealLife \citep{Pivato2007}, SmoothLife \citep{Rafler2011}, Lenia \citep{Chan2019}, and extended Lenia discussed in this paper. They generalize GoL into continuous space using arbitrary long range neighborhoods, into continuous time using arbitrary small incremental updates, and into continuous states using real numbers.

\begin{figure}[hb]
\begin{center}
\includegraphics[width=2.8in]{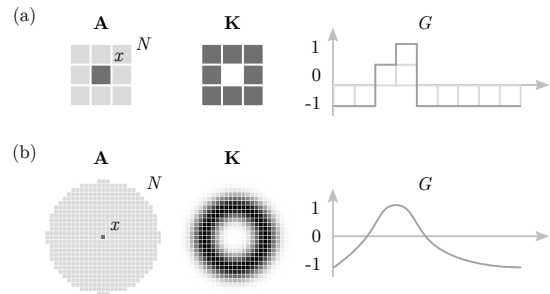}
\caption{Rules of GoL and Lenia. (a) In GoL, a site $x$ in the world $\mathbf{A}$ has 8 surrounding sites as its Moore neighborhood $N$. Calculate the weighted sum of $N$ with kernel $\mathbf{K}$ (all weights 1), apply a mapping function $G$ (survival = 0, birth = +1, death = -1), add the value back to the site $x$ and clip it to 0 or 1, repeat. (b) In Lenia, the rule is similar, but generalized to the continuous domain - infinitesimal sites $x$ with real values, circular neighborhood $N$, ring-like kernel $\mathbf{K}$, smooth mapping $G$, and incremental update by factor $dt$.}
\label{fig1}
\end{center}
\end{figure}

The algorithm of Lenia is as follows (see Figure~\ref{fig1}).
\begin{enumerate}
\item Take a 2D array (world $\mathbf{A}$) of real values between 0 and 1, initialize with an initial pattern $\mathbf{A}^0$.
\item Calculate weighted sums of $\mathbf{A}$ with a predefined array (kernel $\mathbf{K}$), which is equivalent to calculate the convolution $\mathbf{K} \ast \mathbf{A}$; the kernel $\mathbf{K}$ has radius $R$, forming a ring or multiple concentric rings (parameter $\beta$ = list of peak value of each ring).
\item Apply a growth mapping function $G$ to the weighted sums; the growth mapping $G$ is any unimodal function (parameters $\mu$ = growth center, $\sigma$ = growth width).
\item Add a small portion $dt$ of the values back to the array $\mathbf{A}$.
\item Finally clip the states of $\mathbf{A}$ to between 0 and 1.
\item Repeat steps 2-5 for each time-step.
\end{enumerate}

In formula:
\begin{eqnarray}
\mathbf{A}^{t+dt} &= \left[ \mathbf{A}^{t} + dt \; G ( \mathbf{K} * \mathbf{A}^{t} ) \right]_0^1
\end{eqnarray}

In such a continuous CA system, many self-organizing, autonomous solitons were discovered with diverse structures and behaviors. Structures include symmetries like bilateral, radial and rotational symmetries, linear polymerized long-chains, and irregular structures. Behaviors include regular modes of locomotion like stationary, directional, rotating, gyrating, and irregular behaviors like chaotic movements, metamorphosis (shape-shifting), and particle collisions.

The current on-going work is aimed to answer the following open questions raised in the original Lenia paper \citep{Chan2019}:
\begin{enumerate}
\setcounter{enumi}{8}
\item Do self-replicating and pattern-emitting lifeforms exist in Lenia?
\item Do lifeforms exist in other variants of Lenia (e.g. 3D)?
\end{enumerate}
We answer ``Yes'' to both questions. By exploring variants and generalizations of Lenia, we discovered new types of solitons with a wide range of unseen behaviors including self-replication and pattern emission. The current work also aims towards answering Lenia's relationship with Turing completeness (question 6), open-ended evolution (question 7), and other implications in artificial life and artificial intelligence.

\section{Related Works}

SmoothLife \citep{Rafler2011}, an earlier independent discovery similar to Lenia, was the first to report solitons (called ``smooth gliders'') in a continuous 2D CA.

Extensions to Lenia rules were inspired by numerous works about CAs in the literature and in code repositories. There were various attempts in taking existing 2D CAs and other artificial life systems into higher dimensions \citep{Bays1987,Imai2010,Rafler2011b,Sayama2012,Hutton2012b}. Duplication of components in existing CA rules were demonstrated to produce very different dynamics, e.g. Multiple Neighborhoods CA (MNCA) \citep{Rampe2018a,Rampe2018b}, multiple layer CA ``Conway's Ecosystem'' \citep{Sherrill2019}. There were also efforts to blur the boundary between CA and neural networks and brought amazing breakthroughs, e.g. Neural CA \citep{Mordvintsev2020}.

The results of the current work can be compared with other artificial life models, especially particle systems with multiple species of particles, e.g. Swarm Chemistry \citep{Sayama2009}, Primordial Particle Systems \citep{Schmickl2016}, Clusters \citep{Ventrella2017}, developed from the pioneering Boids \citep{Reynolds1987}. These models are able to generate cell-like structures of various styles.

\section{Methods}

Inspired by the related works, we experimented with 3 major extensions to the original Lenia, namely higher dimensions, multiple kernels, multiple channels, and any combinations thereof. We updated the existing open-source software, designed semi-automatic algorithms to search for new patterns and solitons, and performed qualitative analysis on the results.

\subsection{Rule Extensions}

\subsubsection{Higher dimensions}
The 2D arrays in Lenia were upgraded to 3 or higher dimensions, and the algorithms used in the software were subsequently generalized to deal with multidimensional arrays. The number of dimensions is denoted as $d$. Experiments of 3D Lenia have been carried out before but without success in finding interesting patterns. With the utilization of GPU parallel computing and better searching algorithms, stable solitons have been found.

\subsubsection{Multiple kernels}
The original Lenia involves one kernel $\mathbf{K}$ with radius $R$, one growth mapping $G$, and one increment factor $dt$. Now multiply the rule with multiple kernels $\mathbf{K}_k$, each with relative radius $r_k R$, and corresponding growth mapping $G_k$. Weighted average of the results by factors $h_k / h$ ($h$ is the sum of $h_k$) is taken. The number of kernels is denoted as $n_k$. This extension was inspired by MNCA \citep{Rampe2018a,Rampe2018b} that produces highly irregular and dynamic patterns.

\subsubsection{Multiple channels}
Lenia and most CAs have only one world array $\mathbf{A}$, so we experimented with ``parallel worlds'' or multiple channels $\mathbf{A}_i$. In addition to the kernels feeding back to each channel, there are also cross-channel kernels for the channels to interact with each other. Denote the number of channels as $c$, the number of self-interacting kernels per channel as $k_s$, and the number of cross-channel kernels per channel pair as $k_x$, then the total number of kernels $n_k = k_s c + k_x c (c-1)$. This was inspired by multi-layer CA \citep{Sherrill2019} and Neural CA \citep{Mordvintsev2020}.

\subsubsection{Combinations}
The above extensions (and potentially others) can be further combined to produce unique results, e.g. 3D 3-channel 3-self-kernel. The original Lenia becomes a special case, i.e. 2D 1-channel 1-kernel Lenia.

\begin{figure}[hb]
\begin{center}
\includegraphics[width=2.8in]{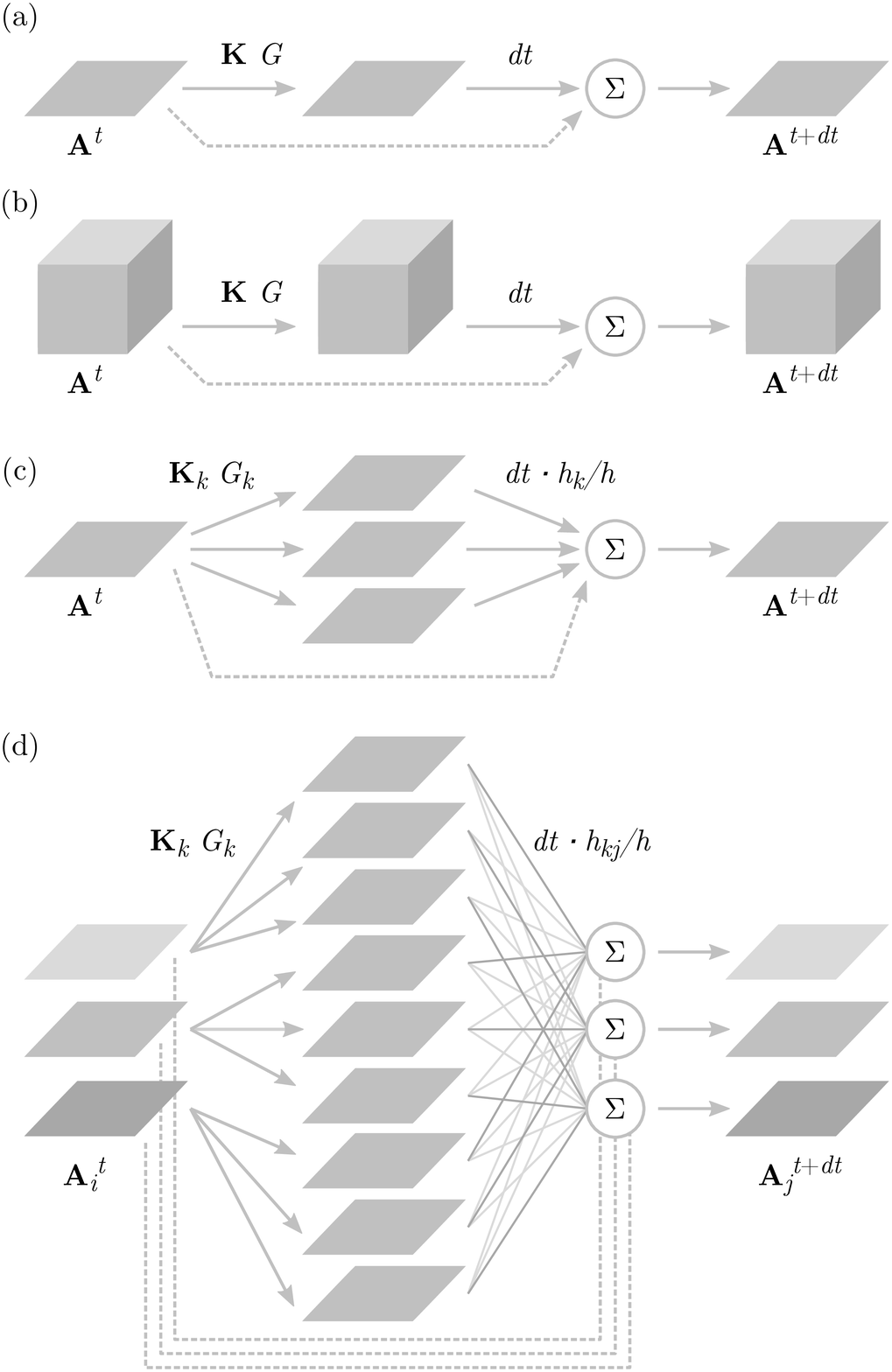}
\caption{Extended Lenia rules. (a) Original 2D Lenia: world $\mathbf{A}$ at time $t$ passes through convolution with kernel $\mathbf{K}$, growth mapping $G$, and incremental update $\Sigma$ to next time step $t+dt$. (b) Higher dimensions with $d$-dimensional arrays. (c) Multiple kernels, where multiple $\mathbf{K}_k$ and $G_k$ feed into $\Sigma$ by factors $h_k$. (d) Multiple channels, where separate channels of world $\mathbf{A}_i$ pass through $\mathbf{K}_k$ and $G_k$, feed into multiple $\Sigma$ that update channel $\mathbf{A}_j$. The architecture approaches a recurrent convolutional neural network.}
\label{fig2}
\end{center}
\end{figure}

The algorithm of extended Lenia is summarized as follows (see Figure~\ref{fig2}).
\begin{enumerate}
\item Create multiple channels of world $\mathbf{A}_i (i=1 \dots c)$, each channel a $d$-dimensional array of real values between 0 and 1; initialize each channel with initial pattern $\mathbf{A}_i^0$.
\item Define multiple $d$-dimensional arrays of kernels $\mathbf{K}_k (k=1 \dots n_k)$, each with relative radius $r_k R$, parameter $\beta_k$, source channel $i$, destination channel $j$, and corresponding growth mapping $G_k$ with parameters $\mu_k$ and $\sigma_k$.
\item For each kernel $\mathbf{K}_k$, calculate weighted sums with its source channel $\mathbf{A}_i$, i.e. convolution $\mathbf{K}_k \ast \mathbf{A}_i$.
\item Apply growth mapping $G_k$ to the weighted sums.
\item Add a small relative portion $dt \cdot h_k / h$ of the values to destination channel $\mathbf{A}_j$.
\item Repeat steps 3-5 for every kernel $\mathbf{K}_k$.
\item Finally clip the states of each channel $\mathbf{A}_i$ to between 0 and 1.
\item Repeat steps 3-7 for each time-step.
\end{enumerate}

In formula:
\begin{eqnarray}
\mathbf{A}_j^{t+dt} &= \left[ \mathbf{A}_j^{t} + dt \sum_{i,k} \frac{h_k}{h} G_k ( \mathbf{K}_k * \mathbf{A}_i^{t} ) \right]_0^1
\end{eqnarray}

\subsection{Genotypes, Phenotypes, and Search Space}

The search space of extended Lenia consists of all possible genotypes and phenotypes. A {\it genotype} here is a particular combination of rule parameter values, a {\it phenotype} is a particular configuration of the world arrays. A pattern (or a soliton) is jointly specified by its genotype and phenotype.

Consider a moderately complex rule of 3D 3-channel 3-self-kernel, with all kernels composed of 3 concentric rings, and a soliton size of $20 \times 20 \times 20$ sites. In this case, the genotype is in the form $(r, h, \beta^3, \mu, \sigma)^{15}$, that is 105 parameter values, and the phenotype consists of 3 channels of 3-dimensional arrays, amounting to 24000 site values.

\subsection{Search Algorithms}

We want to search for interesting patterns or solitons given the new rules. However, the rules create higher degrees of freedom, hence summon the curse of dimensionality. The size of the search space now grows exponentially, manual parameter search and pattern manipulations become difficult if not impossible.  We employed several semi-automatic search algorithms with an interactive user interface to tackle this problem and help exploring the search space.

The algorithms pick genotypes and phenotypes according to some criteria in the search space, and automatically filter them by survival, i.e. to check that the solitons will not come to vanish or occupy the whole grid after running the CA for a period of time. The results are then selected by the human-in-loop for novelty, visual appeal, or prospects for further study, and used in further rounds of semi-automatic search.

\subsubsection{Global search}
The algorithm generates random genotypes and phenotypes from the global search space. The ranges of random values can be tuned to narrow down the search. Once interesting patterns or solitons are found, they can be fed to other algorithms.

\subsubsection{Depth-first search}
Starting with an initial soliton, the algorithm adds small random deviations to one or all values in its genotype, and tests if the phenotype survives. If it does, record the survived phenotype, repeat the process using this new genotype and phenotype as the starting point. This method allows deeper explorations of the search space.

\subsubsection{Breadth-first search}
This algorithm is similar to depth-first search, but using the initial genotype and phenotype as the starting point in every search. This method is able to explore variations of one particular interesting soliton.

\subsubsection{Genetic algorithm}
First set an fitness function and optimization goal (e.g. faster moving speed, higher mass oscillation). Starting from an initial soliton in a pool of samples, the genetic algorithm aggregates the pool using two genetic operators, (1) mutation: pick a random sample from the pool and randomly mutate its genotype; (2) recombination: pick two random samples, create a new sample by randomly mixing their channels and associated parameters. After checking for survival, calculate the fitness value of the new sample, add it to the pool, and sort the pool by fitness. Finally the samples with top fitnesses are recorded as results.

\begin{figure*}[!t]
\begin{center}
\begin{tabular}{@{ }p{0.8in}@{ }p{0.8in}@{ }p{0.8in}@{ }p{0.8in}@{\hspace{0.2in}}p{0.8in}@{ }p{0.8in}@{ }p{0.8in}@{ }p{0.8in}}
1.&2.&3.&4.&1.&2.&3.&4.
\end{tabular}\\
\begin{tabular}{@{ }c@{ }c@{ }c@{ }c@{\hspace{0.2in}}c@{ }c@{ }c@{ }c@{ }}
\includegraphics[width=0.8in]{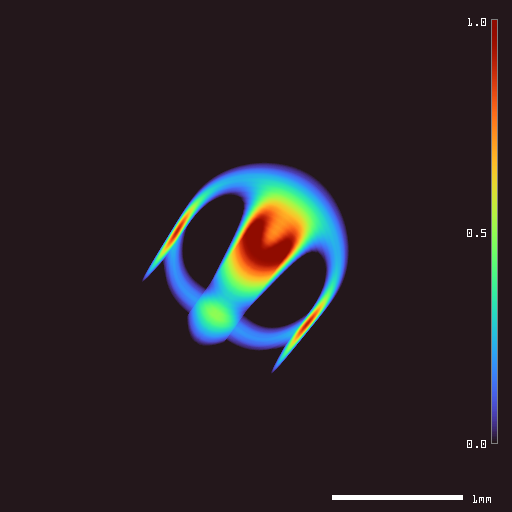} &
\includegraphics[width=0.8in]{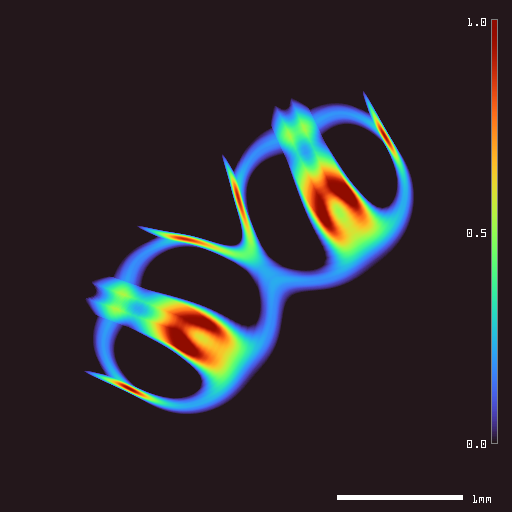} &
\includegraphics[width=0.8in]{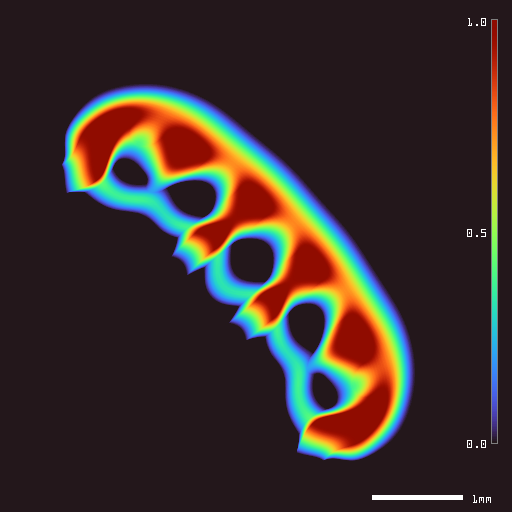} &
\includegraphics[width=0.8in]{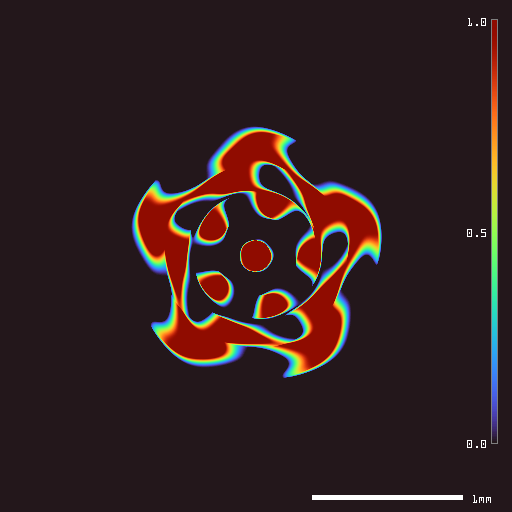} &
\includegraphics[width=0.8in]{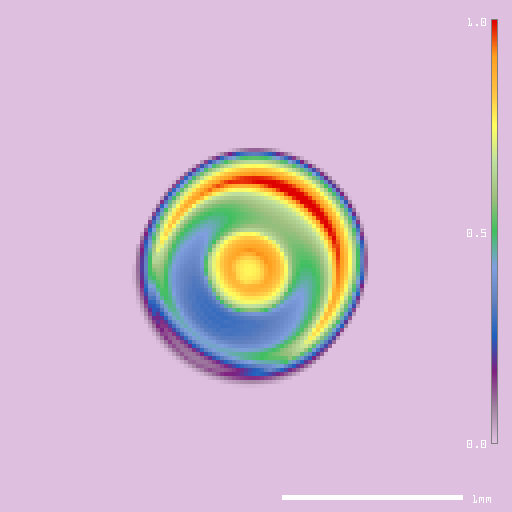} &
\includegraphics[width=0.8in]{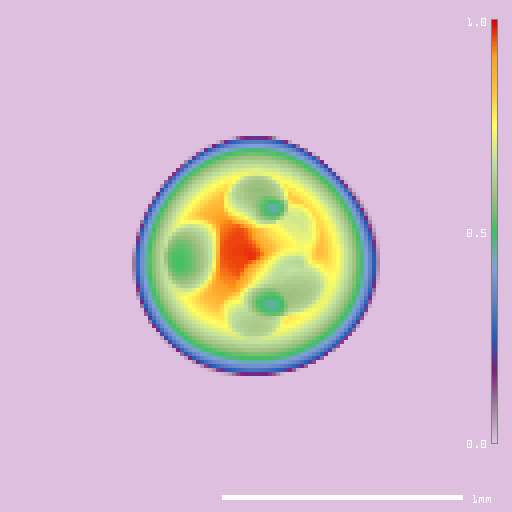} &
\includegraphics[width=0.8in]{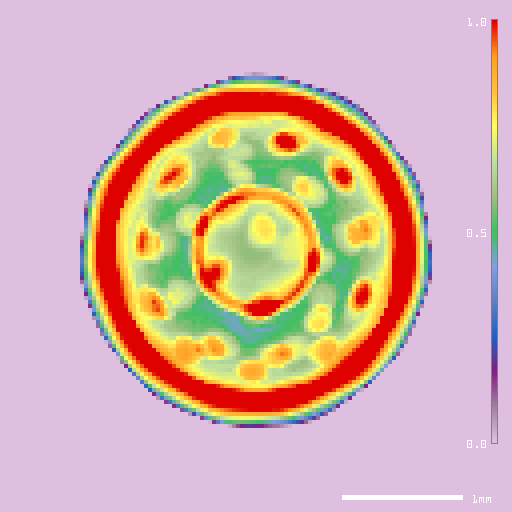} &
\includegraphics[width=0.8in]{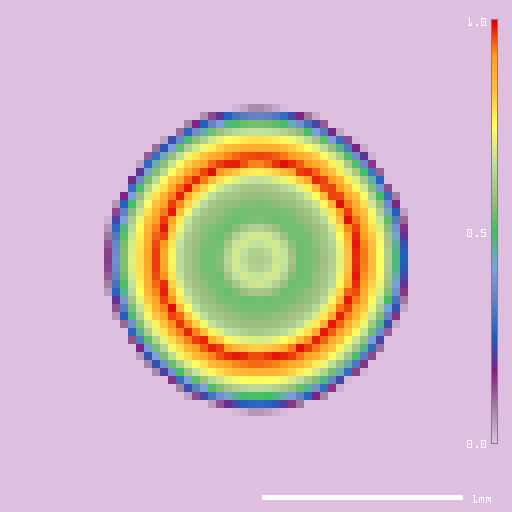} \\
\multicolumn{4}{@{}p{3.3in}}{\small{(a) Original Lenia: 1. {\it Orbium}; 2. {\it Orbium} individuals in elastic collision; 3. long-chain {\it Pentaptera}; 4. rotating {\it Asterium} with 5-fold rotational symmetry.}} &
\multicolumn{4}{@{}p{3.3in}}{\small{(e) Higher dimensions Lenia: 1. moving sphere; 2. rotating sphere with bubbles in trigonal bipyramidal arrangement; 3. pulsating sphere with dots; 4. pulsating 4D hypersphere, showing a 3D slice.}} \\
\\
\includegraphics[width=0.8in]{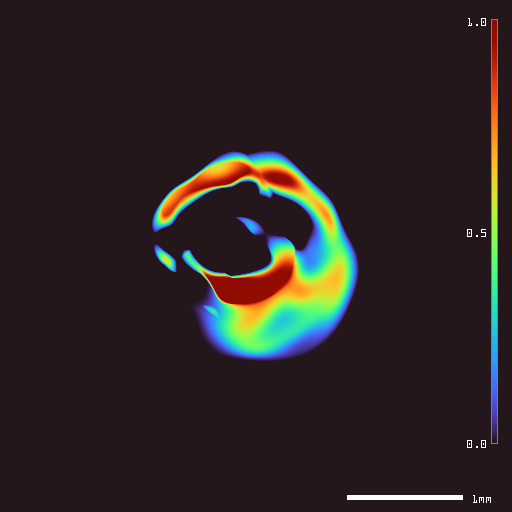} &
\includegraphics[width=0.8in]{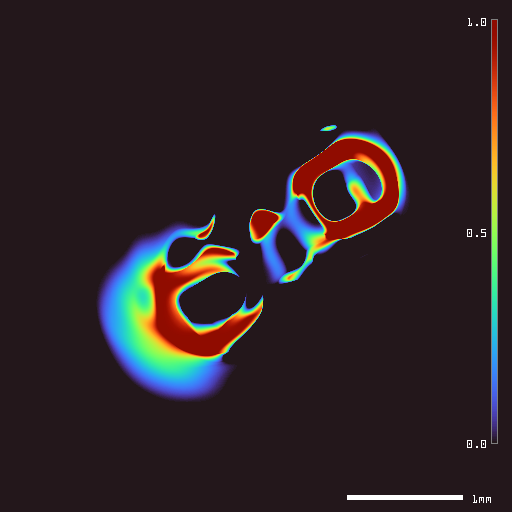} &
\includegraphics[width=0.8in]{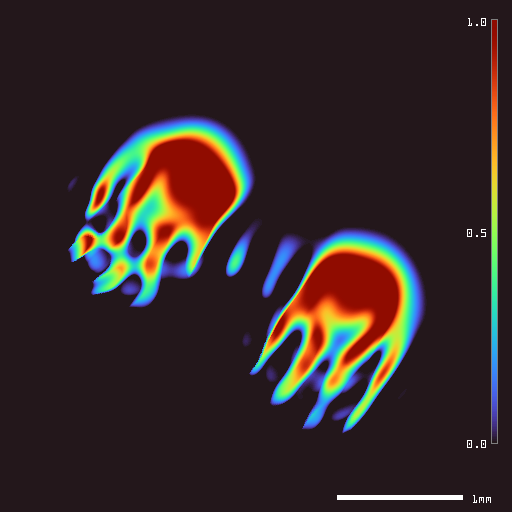} &
\includegraphics[width=0.8in]{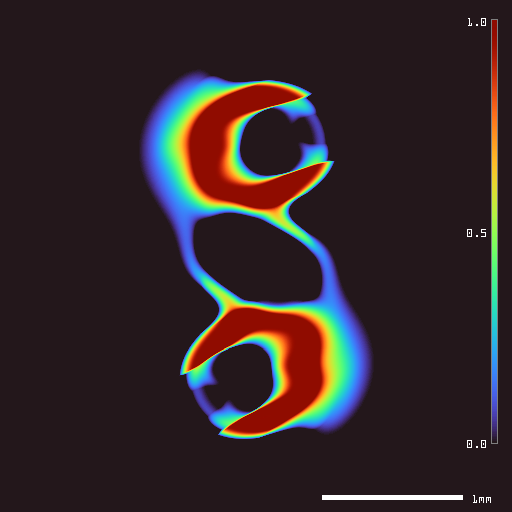} &
\includegraphics[width=0.8in]{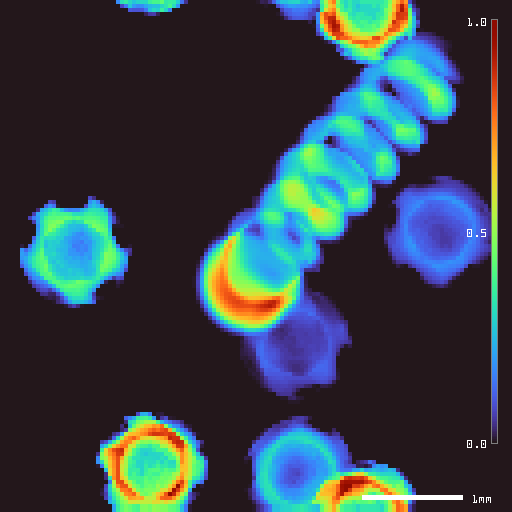} &
\includegraphics[width=0.8in]{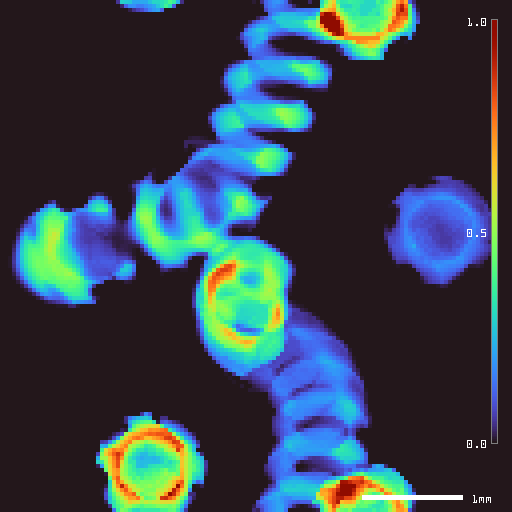} &
\includegraphics[width=0.8in]{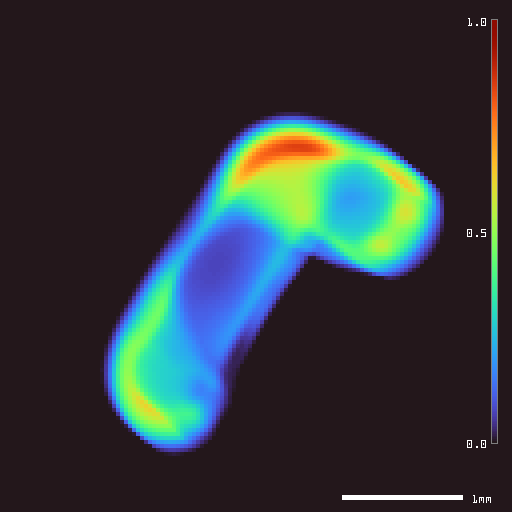} &
\includegraphics[width=0.8in]{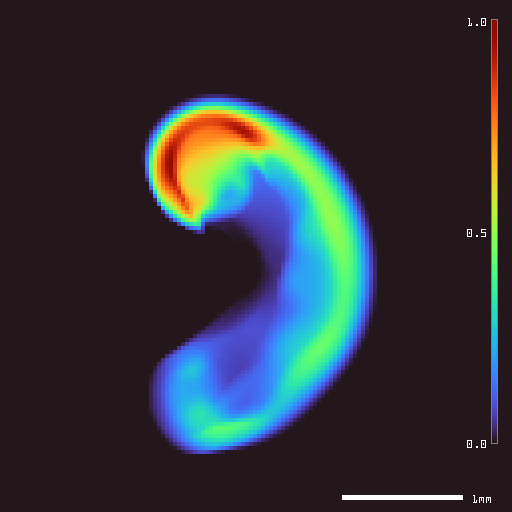} \\
\multicolumn{4}{@{}p{3.3in}}{\small{(b) Multi-kernel Lenia: 1. the first replicator discovered; 2. right after its self-replication; 3. solitons in parallel pair; 4. solitons in elastic collision, repulsive forces hinted by electricity-like lines.}} &
\multicolumn{4}{@{}p{3.3in}}{\small{(f) 3D multi-kernel Lenia: 1. moving ``Snake'' and static ``food dots''; 2. Snake grows while ingesting 3 dots (now spans across the screen); 3-4. a mutant of Snake performing elegant dance.}} \\
\\
\includegraphics[width=0.8in]{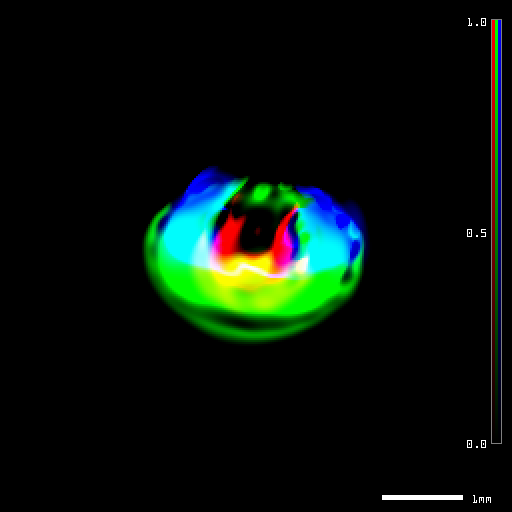} &
\includegraphics[width=0.8in]{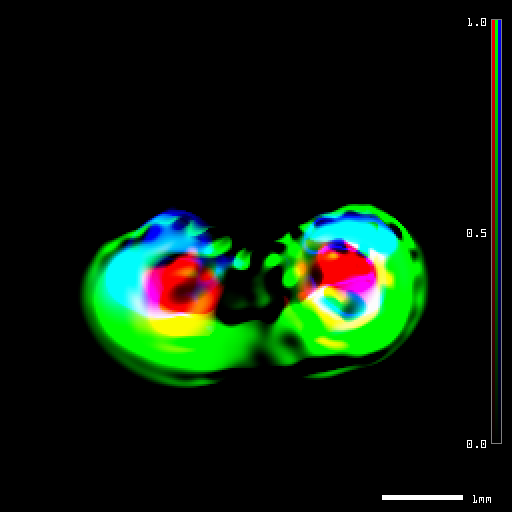} &
\includegraphics[width=0.8in]{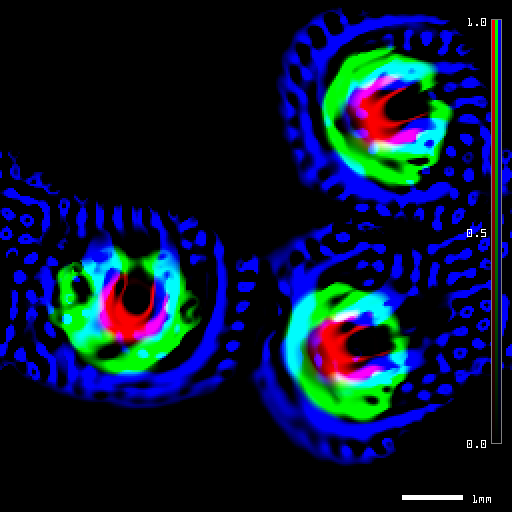} &
\includegraphics[width=0.8in]{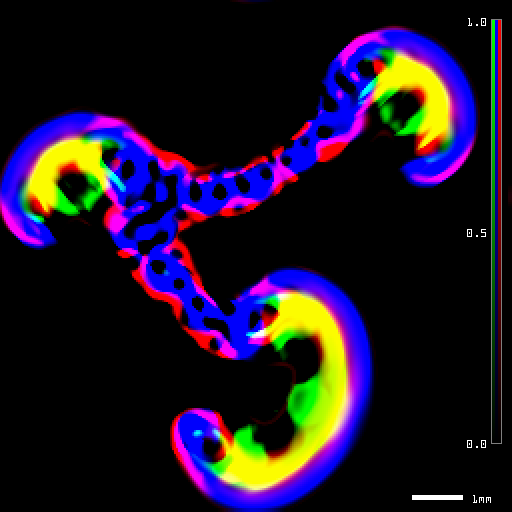} &
\includegraphics[width=0.8in]{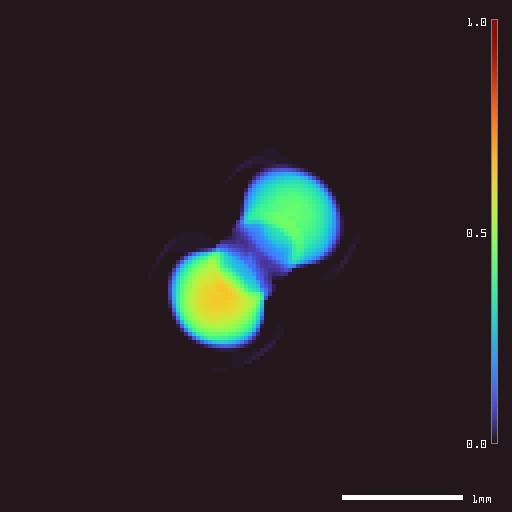} &
\includegraphics[width=0.8in]{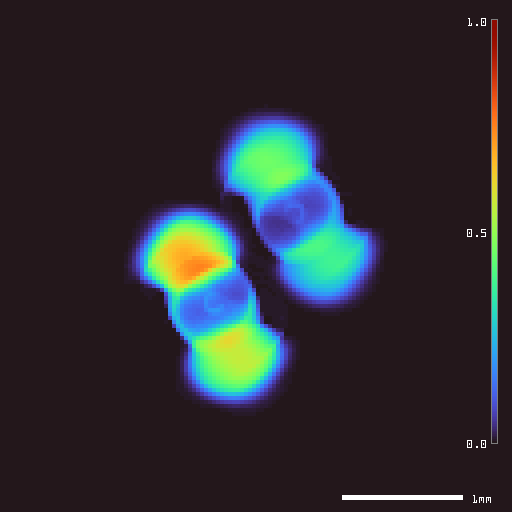} &
\includegraphics[width=0.8in]{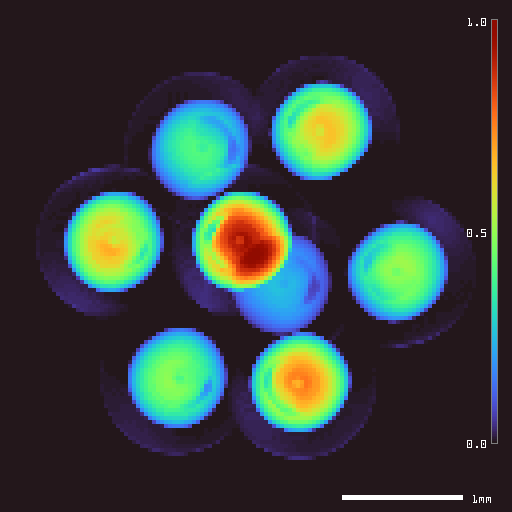} &
\includegraphics[width=0.8in]{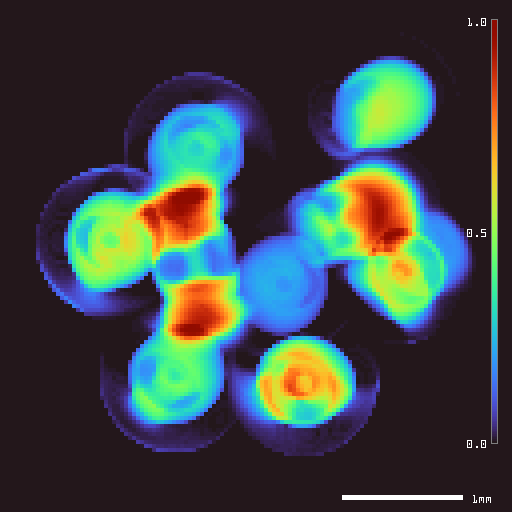} \\
\multicolumn{4}{@{}p{3.3in}}{\small{(c) Multi-channel Lenial: 1. aggregated soliton with cell-like structures; 2. right after its self-replication; 3. sea of emitted particles; 4. dendrite-like emissions from replicating solitons.}} &
\multicolumn{4}{@{}p{3.3in}}{\small{(g) Exponential growth: 1-3. replicator under three rounds of binary fission, repulsive forces visible as negative spheres; 4. Offsprings migrate out for further replication.}} \\
\\
\includegraphics[width=0.8in]{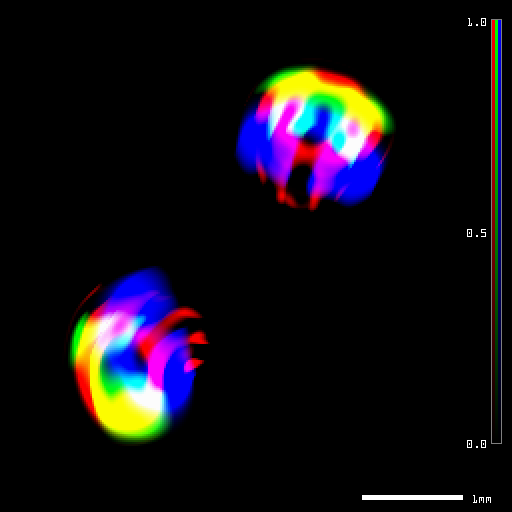} &
\includegraphics[width=0.8in]{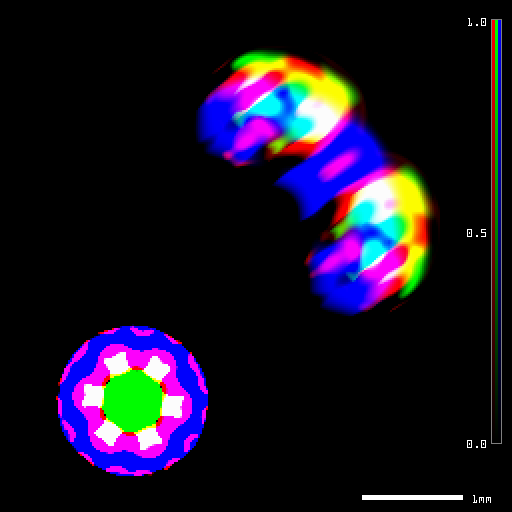} &
\includegraphics[width=0.8in]{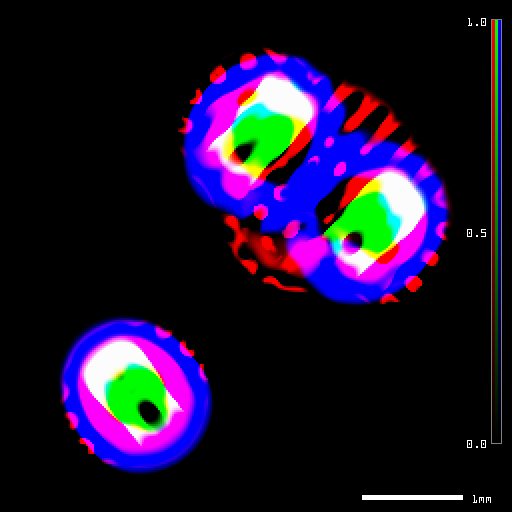} &
\includegraphics[width=0.8in]{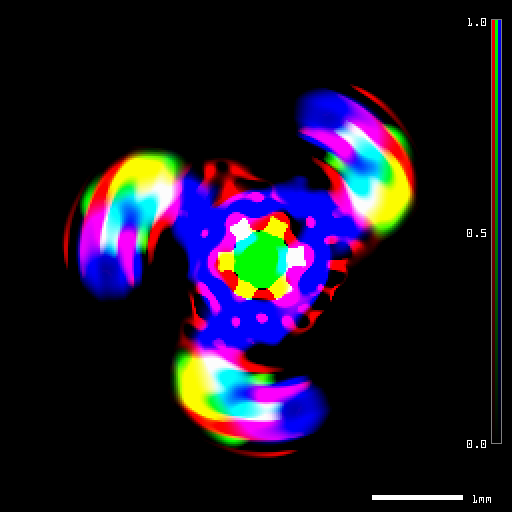} &
\includegraphics[width=0.8in]{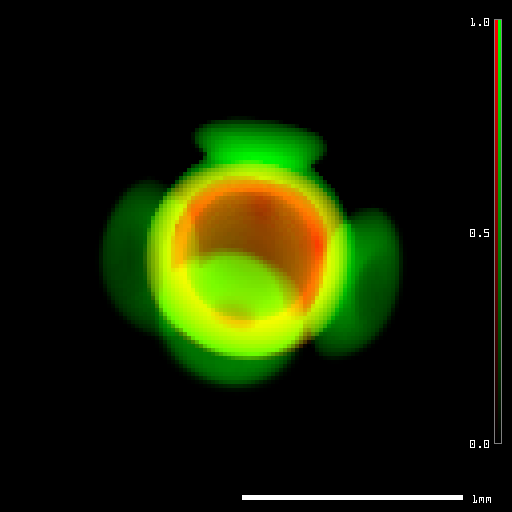} &
\includegraphics[width=0.8in]{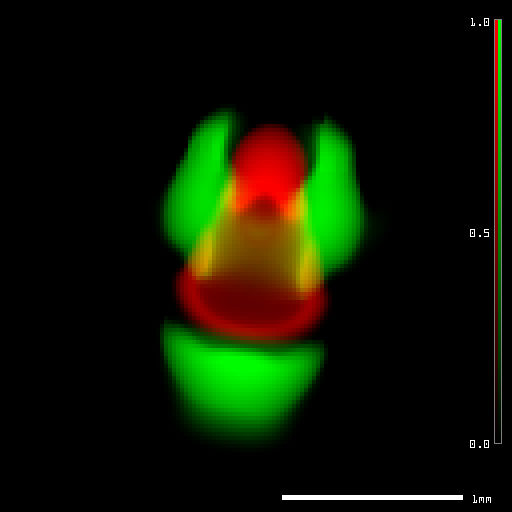} &
\includegraphics[width=0.8in]{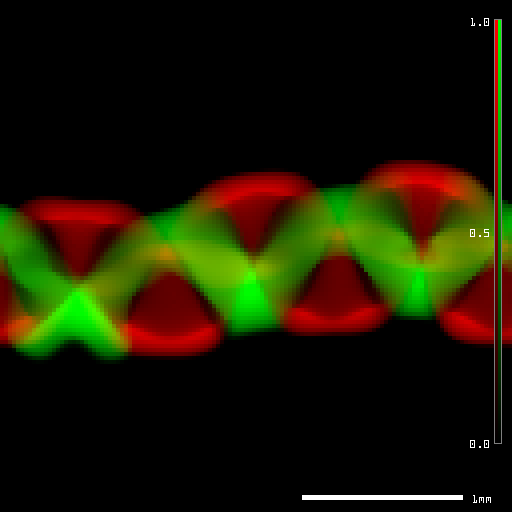} &
\includegraphics[width=0.8in]{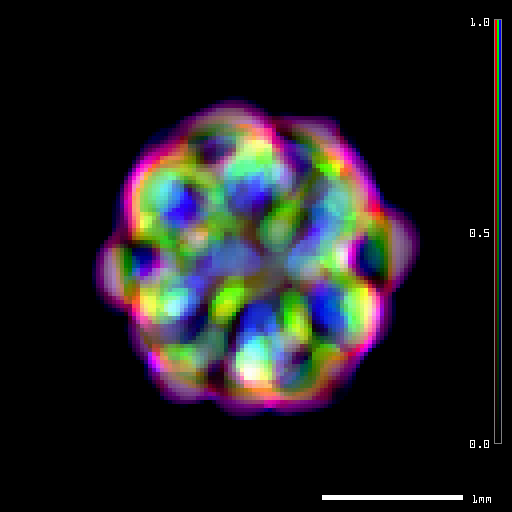} \\
\multicolumn{4}{@{}p{3.3in}}{\small{(d) ``Aquarium'' phenotypes: 1-3. (left to right) gyrating, slightly oblique; stationary, parallel pair; slow-moving, parallel slow-moving; 4. a few solitons in a stable, dynamic formation.}} &
\multicolumn{4}{@{}p{3.3in}}{\small{(h) 3D multi-channel Lenia: 1. tetrapod; 2. moving soliton with red nucleus and green pseudopods; 3. double helix pattern; 4. rainbow ball.}}
\end{tabular}
\end{center}
\caption{Sample solitons. Scale bar at lower right represents kernel radius $R$.}
\label{fig3}
\end{figure*}

\subsection{Software}

The interactive software for Lenia, now open source in GitHub, was updated with the above rule extensions and search algorithms.

For visualization of higher dimensions, the 3D world is flattened to 2D using a depth map, which can show the internal structures of 3D objects with transparency. For dimensions higher than 3, one 3D slice of the array is displayed.

The default color palette used for single-channel visualization was changed from Jet to Turbo \citep{Mikhailov2019} for better perceptual uniformity. For higher dimensions, Paul Tol's Rainbow palette \citep{Tol2018} is recommended to show 3D internal structures. For multiple channels, the first three channels are displayed in red, green and blue (RGB).

\section{Results}

With the help of semi-automatic algorithms, we discovered a number of new structures and behaviors in the extended rules. Unlike the original Lenia, where most solitons are well defined and moderately symmetric, solitons found in the extended rules either possess even higher symmetries (in higher dimensions), or become highly chaotic yet highly self-organized and persistent (with multiple kernels or channels). See Figure~\ref{fig3} for samples (include the original Lenia for reference).

\subsection{Rule Specific Observations}

\subsubsection{Higher dimensions}
In higher dimensions, stable solitons are hard to find, and the found ones are highly stable. Their external shapes are almost always spherical, and their internal structures can be complex and highly symmetrical. In some cases, bubbles (inner voids) are arranged as vertices of Platonic solids or regular polyhedra, e.g. tetrahedron, octahedron, triangular bipyramid, and icosahedron. Most solitons are motionless, a few of them are oscillating, rotating, or directional moving.

Higher dimensional structures are not too chaotic even with multi-kernel or multi-channel extensions, which are supposed to introduce a lot of instability.

\subsubsection{Multiple kernels}
As demonstrated by MNCA, multiple kernels could introduce instability and interesting dynamics into the complex system. Overall chaoticity of the CA increases, but given the right parameters, the system can achieve even higher degrees of self-organization and persistence. There we discovered new or more common behaviors - individuality, self-replication, emission, growth, etc.

\subsubsection{Multiple channels}
In a multi-channel world, each channel develops patterns according to its own rule, and at the same time, these patterns co-develop and influence each other through channel-channel interactions. Different channels of a soliton could exhibit something like a division of labor, e.g. some channels act as outer flexible shells (membranes), some form central masses (nuclei), together they form cell-like structures. In a special case, a particular type of ``Aquarium'' genotype could produce an array of phenotypes, come with different behaviors and complex interactions.

\subsection{Common Phenomena}

\begin{figure}[t]
\begin{center}
\includegraphics[width=2.8in]{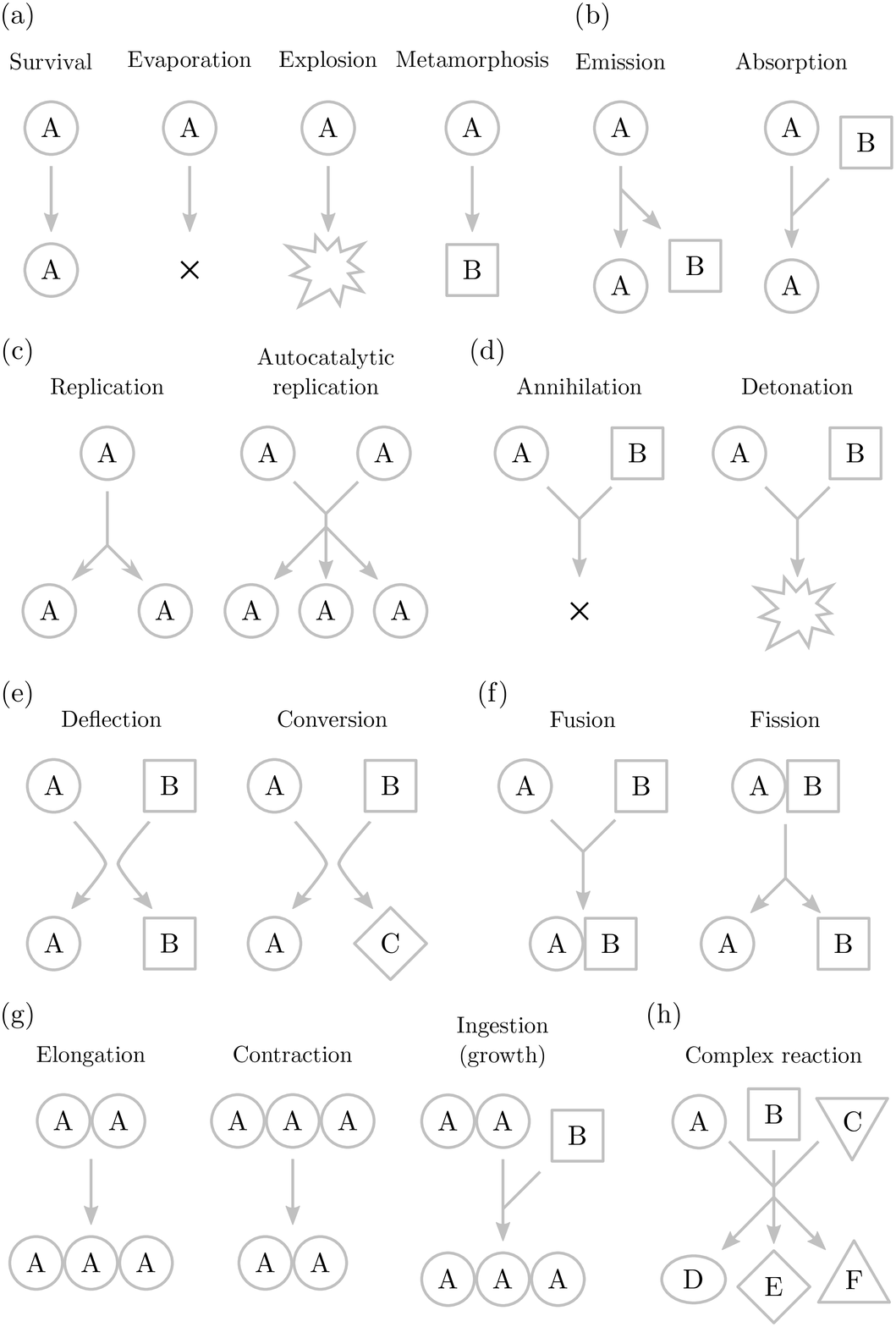}
\caption{Behaviors and interactions of solitons in extended Lenia. Categories: (a) single soliton developments, (b) simple reactions, (c) reproduction, (d) mutual destruction, (e) elastic collisions, (f) inelastic collisions, (g) long-chain reactions, (h) complex reactions.}
\label{fig4}
\end{center}
\end{figure}

We summarize common soliton behaviors and phenomena that can be seen across rules. Refer to Figure~\ref{fig4} for schematic illustrations.

\subsubsection{Locomotion}
In the original Lenia, solitons engage in various kinds of locomotory behaviors, like stationary, directional, rotating, gyrating, oscillating, alternating, drifting, and chaotic movements. In extended Lenia, these movements are still observed, but rotation becomes very rare, possibly because there are fewer cases of rotational symmetry. With multi-kernel and multi-channel, chaotic movements and metamorphosis (shape-shifting) become more prevalent than regular behaviors. Conversely, in 3 or higher dimensions, solitons become predominantly stationary.

\subsubsection{Individuality}
Among the soliton species in the original Lenia, only the Orbidae family (out of 18 families) engages in some forms of elastic or inelastic collisions - when two {\it Orbium} individuals collide, they often reflect each other and survive, or occasionally stick together to form a composite soliton {\it Synorbium}. For other species, solitons in collision simply lose self-organization and die out. Thus {\it Orbium} possesses some kind of {\em individuality}, in that each soliton is able to maintain its own boundary or ``personal space'' and avoid mixing its contents with others.

In multi-kernel or multi-channel rules, {\it Orbium}-like individuality becomes a common phenomenon. Numerous types of solitons manage to maintain self-organization upon collision, thus are able to involve in complex particle interactions. It is possible that some of their kernels or channels act as repelling forces that separate individuals from each other.

\subsubsection{Self-replication}
An important milestone in the study of Lenia is the discovery of {\em self-replication}. It is conspicuously missing in the original Lenia, but turns out to be not rare in extended rules. The mechanism is usually one soliton develops into two partitions of similar structures, each develops into a full soliton, drifts away, and is capable of further division. In highly reproductive cases, new individuals can develop out of debris. In multi-channel rule, self-replication is usually initiated by division in one channel, then other channels follow suit. Self-replication is closely related to individuality - newly replicated parts need to repel and separate from each other to complete the process.

There is also {\em autocatalytic replication}. In some cases, self-replication does not or only seldom happens when the density of solitons is low. But when the density rises (e.g. from the very slow reproduction), congregation of solitons will force self-replication to happen, kicks start a wave of autocatalysis and causes exponential growth.

Reproducing solitons occupy all available space sooner or later. But if those solitons also vanish with a death rate not far from the birth rate, it may maintain a ``healthy'' population of regenerating solitons.

\subsubsection{Growth by ingestion}
We found this curious phenomenon only in one setting (the ``3D Snake'' genotype) of 3D multi-kernel rule. In the Snake world, there is one type of static spherical solitons, ``food dots'', and one type of dynamic helical solitons, ``snakes''. A snake keeps contracting or extending linearly at one or both ends, giving an illusion of a moving snake. When its extending end reaches one food dot, it merges with that ``inanimate'' dot ({\em ingestion}), turns it into part of the ``living'' soliton, and slightly elongates ({\em growth}). The snake also slightly changes direction towards dots within reach, giving an illusion of the snake pursuing food. \footnote{Upon seeing in action, one may be reminded of the ``Snake'' mini-game in Nokia mobile phones, except that the Snake world here is not pre-programmed and snake control is not provided.}

This growth behavior may be related to the elongation and contraction of long-chain species (Pterifera) in the original Lenia. It is probably an exceptional and isolated case, but remarkable that it is even possible to happen.

\subsubsection{Emission}
In GoL, an important category of patterns that enables universal computation is the ``guns'' - stationary patterns that emit moving solitons. There are other categories: ``puffer trains'' (moving emit stationary), ``rakes'' (moving emit moving), and complex tertiary emissions. {\em Pattern emission} is sometimes found in extended Lenia, but is usually irregular and of the ``puffer train'' type. We aim to find more regular, reliable emitters in Lenia, especially of the ``gun'' type, in order to pursue Turing completeness \citep{Berlekamp2018}, or some kind of analog computation.

\subsubsection{Division of labor}
In multi-kernel and multi-channel rules, various channels and kernels engage in different behaviors yet influence each other. As discussed above, some kernels or channels may form patterns that exert repulsion and define the scope of the pattern, some may facilitate binary fission, some engage in pattern emission; some may provide stability and some others provide motility.

Dynamic or static patterns from different channels combine into an {\em aggregated soliton}. For the aggregated soliton to survive and prosper, its channels must coordinate and cooperate with each other. It acts as a single unit, engages in diverse complex behaviors, and evolves as a whole.

\subsubsection{Differentiation}
We found a special range of ``Aquarium'' genotypes in multi-channel rule, where one genotype produces multiple phenotypes of aggregated solitons, each having own stable structure and behavior.

The collection may include solitons with directional ({\it rectus}), oblique ({\it limus}), gyrating ({\it gyrans}), stationary ({\it lithos}), slower or faster moving ({\it tardus} or {\it tachus}), parallel / antiparallel pairing ({\it para-} / {\it anti-}) phenotypes, and possibly more. Each of the phenotypes is usually quite stable and well defined, but can switch to another phenotype in specific occasions, e.g. upon collision or after self-replication.

This is a desirable emergent property in Lenia, since it enables heterogeneous soliton-soliton interactions for the first time. Complex interactions and reactions, together with self-replication, may lead to higher-level structures and collective behaviors, like building up tissue-like megastructures.

\section{Discussion}

\subsection{Relations to Biology}

The original Lenia, and other models like SmoothLife \citep{Rafler2011}, have shown that continuous CAs are able to produce patterns with appearance and dynamics comparable to real world biology. With more discoveries in extended Lenia, we can add more comparisons between artificial life and biological life.

\subsubsection{Origin of Life}
The gradual emergence of several important phenomena in Lenia is reminiscent of the origin of life.

Cell individuality and self-replication are among the hallmarks of life on Earth, each has abiotic origins. Individuality originated from lipid membranes that were formed spontaneously by hydrophobic molecules in the primordial soup, separate the outside world from an area where specific chemical reactions can occur, and protect such an area from physical attacks and chemical insults \citep{Haldane1929}. Self-replication possibly came from the RNA World, where RNA molecules self-assemble and self-replicate out from amino acid building blocks \citep{Joyce1989}.

Division of labor inside eukaryotic cells, i.e. the cells of all animals, plants and fungi, stemmed from endosymbiosis of more basic lifeforms, i.e. bacteria, archaea, and possibly viruses \citep{Mereschkowsky1905,Sagan1967}. Mitochondria originated from an ancient unification of $\alpha$-proteobacteria with archaea. The bacteria provided aerobic energy metabolism, and the archaea provided the cytoplasm and membrane. Chloroplasts originated from further endosymbiosis with cyanobacteria, equipped algae and plant cells with photosynthesis. The nuclei of the eukaryotic cell may have originated from DNA viruses \citep{Bell2001}. These organelles, together with the cell body, perform various functions separately and also cooperate closely.

Here in extended Lenia, similar processes of individuality, self-replication, and division of labor have emerged from the more and more generalized CA rules.  Is it possible that these processes, and maybe others, are essential in creating more and more complex evolvable systems in both the real world and the virtual world.

\subsubsection{Organization hierarchy}
If we compare the levels of organization in Lenia to the hierarchy of biological structures - from atoms to organisms to ecosystems, we could come up with more than one interpretations (Table~\ref{tab1}). 

\hyphenation{organelles}
\hyphenation{reaction}
\hyphenation{multi-cellular}

\begin{table}[t]
\center{
\begin{tabular}{>{\hangindent=1em}p{1in}>{\hangindent=1em}p{0.95in}>{\hangindent=1em}p{0.95in}}\hline
Lenia & Cellular level & Molecular level\\ \hline
Site & Cell & Molecule\\
Kernel & Cell signaling & Chemical reaction\\
Single-channel soliton & Simple multicellular life & Prokaryote, virus\\
Multi-channel soliton & Complex multicellular life & Eukaryotic cell\\ \hline
Division of labor & Organs & Organelles\\
Center & Heart / brain & Nucleus\\
Individuality & Body, skin & Cytoplasm, membrane\\
Motility & Limb & Pseudopod\\
Emission & Signal & Cytokine\\
Differentiation & Polymorphism & Cell type\\
\hline
\end{tabular}
}
\vskip 0.25cm
\caption{Comparisons of self-organization levels in Lenia to biology.}
\label{tab1}
\end{table}

The straightforward take, as implied in the name ``cellular automata'', is to interpret a site in CA as a biological ``cell'' (or a ``concentration of cells'' in continuous CAs). A neighborhood or kernel would be something like a cell signaling pathway, affecting surrounding cells with a certain effect. In this analogy, single-channel solitons are like simple multicellular organisms without organs (e.g. sponges, jellyfish, fungi, kelps, slime molds), and multi-channel solitons are like complex multicellular organisms (e.g. bilaterian animals, higher plants), with division of labor among organs.

In a more interesting interpretation, a site can be thought of as a ``molecule'' (or a ``concentration of molecules'' in continuous case). Consequently a kernel would be a type of molecular force or chemical reaction, influencing surrounding molecules according to distance and concentration. Single-channel solitons, including those in the original Lenia, would resemble simple microscopic lifeforms (e.g. bacteria, archaea, viruses), possess self-organization, self-replication, symmetry, individuality, motility, etc. Multi-channel solitons, especially of the ``Aquarium'' genotypes, would resemble eukaryotic cells, with internal division of labor among organelles, and differentiation among cell types. 

\begin{figure}[t]
\begin{center}
\includegraphics[width=2.8in]{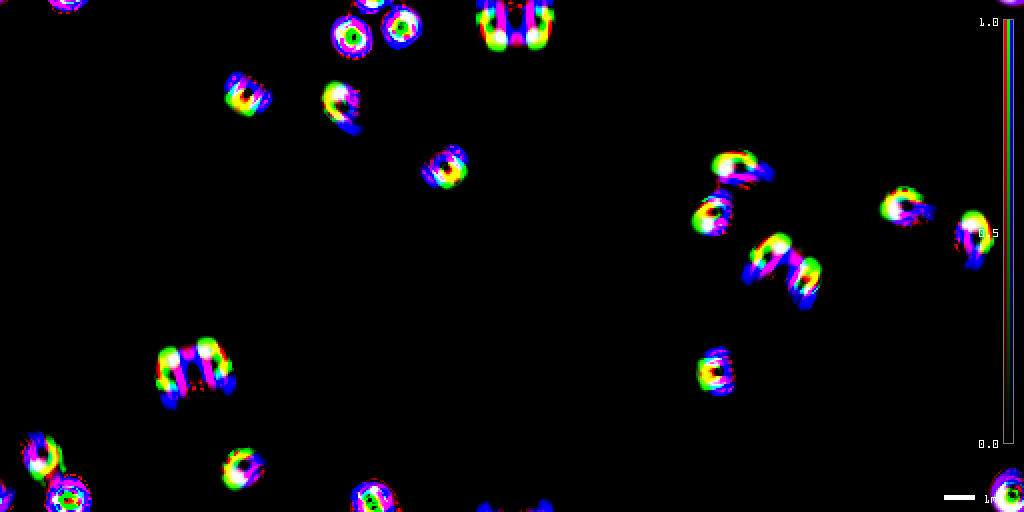}\\
\small{(a)}\\
\includegraphics[width=2.8in]{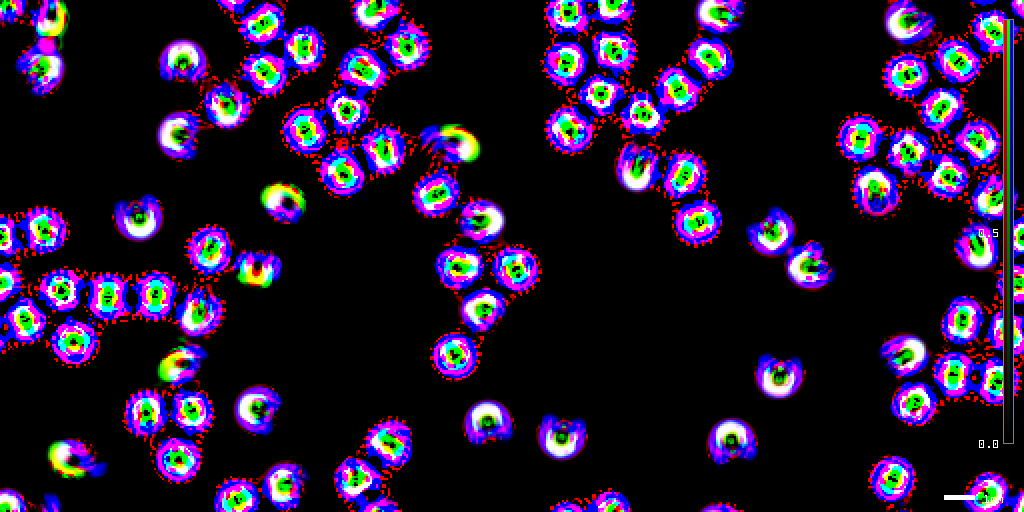}\\
\small{(b)}\\
\caption{``Virtual eukaryotes'' in action. (a) Solitons of ``Aquarium'' set similar to Figure~\ref{fig3}(d), but with a highly reproductive gyrating phenotype, start to reproduce, differentiate, migrate, interact and react with each other. (b) A few tissue-like colonies gradually formed, akin to what happens in multicellularity.}
\label{fig5}
\end{center}
\end{figure}

\subsubsection{Virtual cells}
These multi-channel solitons no longer need different genotypes to realize different behaviors, all they need are subtle changes in the division of labor and coordination of internal parts, express themselves as different phenotypes. The kinds of division of labor observed include:
\begin{itemize}
\item Some channels form a pattern like a ``nucleus'', usually at the center of an entity. Other channels develop patterns around the nucleus. Whenever the nucleus moves, self-replicates, or dies out, other channels usually follow suit.
\item Some channels form ``cytoplasm'' or ``membrane'' that defines a private area around the nucleus, keeps safe distances from other patterns by means of repulsive and attractive forces.
\item Some channels may form movable parts like ``pseudopods'', direct the movement of whole soliton when the pseudopod is at the periphery, or stay stationary when it is kept inside the cytoplasm.
\item Some channels may form ``tails'' behind the soliton (perhaps not for propulsion).
\item Some channels may emit signal-like small particles like ``cytokines'', significance uncertain.
\end{itemize}

In this regard, these complex solitons could be dubbed ``virtual eukaryotes'' or  ``virtual stem cells'' (Figure~\ref{fig5}). They are by far the most lifelike patterns in the Lenia family of continuous CAs.

Altogether, a community of ``virtual eukaryotes'' engages in diverse emergent behaviors and complex interactions thanks to their own high level of self-organization, and it is not impossible that they will later be shown to produce another level of emergence and self-organization.

\subsection{Relations to Other Systems in Artificial Life}

Particle systems (PS), like Swarm Chemistry \citep{Sayama2009}, Primordial Particle Systems \citep{Schmickl2016}, Clusters \citep{Ventrella2017}, have multiple species of particles engage in intra- and inter-species interactions. They produce results that are comparable to multi-channel Lenia. The particles in PSs self-organize into aggregated patterns (solitons), build cell-like structures like cytoplasms, membranes and nuclei, and engage in binary fission, etc. One difference is that solitons in these PSs do not possess strong individuality, hence almost always merge upon collision.

It may be difficult to compare CAs and PSs because of a few fundamental differences in their rulesets - PSs calculate the vector movements of every particle, and maintain a conservation of mass, while CAs only keep track of scalar states and the total mass is not conserved.  To deal with this discrepancy, one may interpret the scalar states in CAs as concentrations of virtual molecules across a grid (see Molecular level column in Table~\ref{tab1}), and the molecules can be constructed, destroyed or migrated with rates according to the CA rule. The relationship between CAs and PSs would be like that of the macroscopic view of thermodynamics vs the microscopic view of Newtonian physics.

\subsection{Relations to Artificial Intelligence}

There are efforts to employ methodologies from artificial intelligence to search for new artificial life patterns. \citet{Reinke2019} used curiosity-based algorithm IMGEP \citep{BaranesOudeyer2013} and neural networks like CPPN and VAE to explore the search space of the original Lenia, with success in increasing the diversity in pattern search. Interactive evolutionary computation (IEC) \citep{Takagi2001} and genetic algorithms (GA) were also used in semi-automatic discovery of new patterns \citep{Chan2019}.

On the other hand, a number of researchers have noticed the close relation between CAs and neural networks (NN) \citep{Wulff1992,Gilpin2018}. \citet{Mordvintsev2020} designed Neural CA, a CA-NN hybrid that can be trained to generate and regenerate (also playfully interpolate) predefined patterns. They suggested that the Neural CA could be named ``Recurrent Residual Convolutional Networks with `per-pixel' Dropout''.

The architecture of our multi-channel Lenia also approaches a ``Recurrent Residual Convolutional Network'' (see Figure~\ref{fig2}(d)). The ``recurrent'', ``convolutional'', and ``residual'' attributes come from the repetitive updates, the convolution kernels, and the contributions from world states, respectively. The growth mapping is analogous to an activation function. The incremental update part vaguely resembles a fully connected layer in NN.

\subsubsection{Comparing Lenia and Neural CA}
Lenia relies on tuning the parameters of kernels and growth mappings to ``train'' the model into generating self-organizing patterns, while the incremental update part has limited flexibility. Neural CA, on the other hand, is fixed in the convolutional kernels and activation functions, but heavily parameterized in the fully connected layers. Lenia is aimed at exploring novel patterns, helped by evolutionary, genetic and exploratory algorithms; Neural CA is aimed at generating predefined patterns, results are optimized by gradient descent.

Despite the differences, Lenia and Neural CA do one thing in common - exploit the self-organizing, emergence-inducing, and regenerating powers of CAs. Neural CA also exploits the learnable nature of its NN architecture, and it remains unknown whether the Lenia model can be made learnable to achieve other goals.

\subsection{Future Works}

The following future works are proposed:
\begin{itemize}
\item Automatic identify and count soliton individuals. This would allow the software to detect individuality, self-replication, birth rate and death rate, soliton interactions, etc., and hence select for these attributes using genetic algorithms.
\item Using ``virtual eukaryotes'' as elements, study the possibility of the next level of emergence and self-organization, and compare the results to multicellularity, cell differentiation, cell signaling in biology.
\item Develop Lenia into trainable Recurrent Residual Convolutional Networks or GANs for whatever purpose.
\end{itemize}

\section{Supplementary Info}

The open-source software of Lenia in Python is available at:
\url{https://github.com/Chakazul/Lenia}

\section{Acknowledgements}

This work is dedicated to the late John H. Conway, inventor of the Game of Life, and the late Richard K. Guy, discoverer of the ``glider'', the first soliton in GoL.

I would like to thank Pierre-Yves Oudeyer and the Inria Flowers team Chris Reinke, Mayalen Etcheverry, Clement Moulin-Frier for intellectual exchanges; Will Cavendish, Cl\'ement Hongler, Gloria Capano, Takaya Arita, Nick Kyparissas, Michael Simkin, Michael Klachko, John Sherrill, Alex Mordvintsev, Craig Reynolds for valuable discussions and inspirations; Hector Zenil, Josh Bongard, Dennis Allison for opportunities in publications and university talk; David Ha, Lana Sinapayen, Sam Kriegman for continued supports in my road as an independent researcher.

\footnotesize
\Urlmuskip=0mu plus 1mu
\bibliographystyle{apalike}
\bibliography{Lenia-Expanded}

\end{document}